\documentclass{article}

\usepackage{arxiv}
\usepackage[utf8]{inputenc} 
\usepackage[T1]{fontenc}    
\usepackage{hyperref}       
\usepackage{url}            
\usepackage{booktabs}       
\usepackage{amsfonts}       
\usepackage{nicefrac}       
\usepackage{microtype}      
\usepackage{lipsum}
\usepackage{graphicx}
\usepackage{amsmath}
\usepackage{amssymb}
\graphicspath{ {./images/} }

\title{Exploiting nonlinear incoherent image formation\\ through linear volume metaoptics \\for inference}

\author{
	Nan Zhang\\
	Bradley Department of Electrical and Computer Engineering\\
	Virginia Tech\\
	Blacksburg, VA 24061 \\
	\texttt{nan25@vt.edu} \\
	\And
	Arvin Keshvari \\
	Bradley Department of Electrical and Computer Engineering\\
	Virginia Tech\\
	Blacksburg, VA 24061 \\
	\And
	Ata Shakeri \\
	Bradley Department of Electrical and Computer Engineering\\
	Virginia Tech\\
	Blacksburg, VA 24061 \\
	\And
	Zin Lin \\
	Bradley Department of Electrical and Computer Engineering\\
	Virginia Tech\\
	Blacksburg, VA 24061 \\
	\texttt{zinlin@vt.edu}\\
}

\begin{document}
	\maketitle
		
\begin{abstract}
	We showed that a 2D depth map representing an incoherent 3D opaque scene is directly encoded in the response function of an imaging optics. As a result, the optics creates an image that depends nonlinearly on the depth map. Furthermore, strong spatio-spectral dispersions in volume metaoptics can be engineered to create a complex image in response to a depth map. We hypothesize that this complexity will allow the linear volume metaoptics to nonlinearly sense and process 3D opaque scenes.
\end{abstract}
	
\section{Introduction}
While weak nonlinearities have long hampered the development of all-optical information processors, recent pioneering works showed that it is possible to perform nonlinear computation by linear optics via clever information-encoding schemes~\cite{wanjura2024fully,eliezer2023tunable,xia2024nonlinear}. So far, these schemes operate on artificially encoded input information---which is often synthesized in the electronic domain---and are primarily geared towards coherent and/or integrated photonics settings. As such, these approaches are not yet applicable to free-space computer-vision or inference tasks which must directly sense and process \textit{incoherent} signals from natural-light scenes. In this short \textit{conceptual} article, we theoretically investigate the feasibility of a class of inference problems where naturally-occurring real-world information, carried by incoherent waves, may be nonlinearly processed by linear volume metaoptics. The latter generalizes local, single-layer metasurface optics~\cite{chen2020flat} to arbitrary non-periodic volumetric nanophotonic structures~\cite{lin2018topology,lin2021computational,roques2022toward,lin2021end,roberts20233d}, offering vastly richer degrees of freedom as well as strongly enhanced spectro-angular dispersions~\cite{lin2018topology,lin2021computational}, which will be crucial to sensitively capturing and arbitrarily processing real-world scene information. 

The key to our approach is the realization that a 3D opaque scene, which can be naturally characterized by a 2D depth map, creates an image which \textit{nonlinearly} depends on the depth map---in contrast to the image of a 3D point cloud, which depends linearly on the point-cloud intensities. This is because the depth map is directly encoded into the response function of the imaging metaoptics. Therefore, a highly depth-sensitive meta-structure, such as a freeform volume metaoptics, may produce complex electric fields and images in response to a depth map. We hypothesize that this complexity will allow linear volume metaoptics to perform nonlinear operations on opaque scenes for deep learning of certain inference tasks. In Section~\ref{sec:imgform}, we delineate the principle of nonlinear image formation from a 2D depth map. In Section~\ref{sec:results}, we report proof-of-concept metaoptics designs which exploit the nonlinear image formation process to perform sophisticated inference tasks on depth maps. In Section~\ref{sec:outlook}, we discuss potential generalizations to more complex light-matter interactions and broader computer vision problems.

\section{Nonlinear Incoherent Image Formation}
\label{sec:imgform}
A real-world 3D scene under incoherent ambient light can be typically represented by a spectral-3D intensity function $u(x,y,z; \lambda)$. By the principle of incoherent image formation, the image $v$ of $u$ is a sum over field intensities generated by the imaging optics in response to each point in the support of $u$ so that
\begin{align}
	v(x,y) = \int G(x,y,z_\text{CCD}; x',y',z'; \lambda)~u(x',y',z') ~ dx' dy' dz' d\lambda \label{eq:imgform}
\end{align}
Here, $G(x,y,z_\text{CCD}; x',y',z';\lambda)$ is the \textit{intensity} response function, which fully characterizes the imaging optics, and is simply the intensity image at the CCD sensor plane ($z=z_\text{CCD}$) of a time-harmonic dipole positioned at $(x',y',z')$ and oscillating at a frequency $\omega = 2\pi c/ \lambda$. Computationally, $G$ can be obtained by simulating the propagation and scattering of electromagnetic waves through the imaging metaoptics, the latter characterized by a spatial permittivity profile $\varepsilon(x,y,z)$:
\begin{align}
	\nabla \times \nabla \times E(x,y,z) - \omega^2 \varepsilon(x,y,z)E(x,y,z) &= i \omega~ \delta(x-x',y-y',z-z'), \label{eq:maxwell} \\
	G(x,y,z_\text{CCD}; x',y',z';\lambda, \varepsilon) &= |E(x,y,z_\text{CCD})|^2.
\end{align}
Note that Eq.~\eqref{eq:maxwell} must be solved repeatedly for all the $(x',y',z')$ points in the support of $u$ as well as for all the wavelengths $\lambda$ in the operational bandwidth of interest. Clearly, in Eq.~\eqref{eq:imgform}, $v$ depends linearly on $u$ so that it cannot be exploited to perform nonlinear inferential computations on $u$, no matter how complex $G$ is. 

If the 3D scene is opaque, the intensity representation $u(x,y,z; \lambda)$ is highly sparse and can be equivalently reduced to a spectral-2D intensity map $u_\text{2D}(x,y;\lambda)$ and a 2D depth map $h(x,y)$. Mathematically, an opaque scene means that at each $(x,y)$, there is at most only one $z$ coordinate where the intensity is non-zero. We capture this insight with a delta function:
\begin{align}
	u(x,y,z; \lambda) = u_\text{2D}(x,y; \lambda) \delta( z - h(x,y) ) \label{eq:3Ddepth}
\end{align}
Substituting Eq.~\eqref{eq:3Ddepth} into Eq.~\eqref{eq:imgform}, we obtain:
\begin{align}
	v(x,y) &= \int G(x,y,z_\text{CCD}; x',y',z'; \lambda, \varepsilon)~u_\text{2D}(x',y'; \lambda) \delta( z' - h(x',y') ) ~ dx' dy' dz' d\lambda \\
	&= \int G(x,y,z_\text{CCD}; x',y',h(x',y'); \lambda, \varepsilon)~u_\text{2D}(x',y'; \lambda) ~ dx' dy' d\lambda \label{eq:NLimgform}
\end{align}
Crucially, Eq.~\eqref{eq:NLimgform} indicates that the real-world information $h$ is encoded into the response function $G$ of the imaging optics itself. Since $G$ depends nonlinearly on $h$, so does $v$. In a more succinct notation, the image $v$ is a nonlinear function $f$ of the input $h$ as well as the ``trainable'' parameters $\varepsilon$:
\begin{align}
	v = f(h; \varepsilon) \label{eq:NLvf}
\end{align}
The function $f$ may be highly complex if the permittivity $\varepsilon$ is sufficiently complex (as in volume metaoptics). Thus, Eq.~\eqref{eq:NLvf} provides a potentially \textit{expressive} nonlinear model---though not without limitations---where $\varepsilon$ can be trained (i.e., freeform-inverse-designed~\cite{molesky2018inverse}) via adjoint optimization to perform nonlinear operations and inference on real-world information $h$. In practice, we may also include an additive Gaussian noise term $v = f(h; \varepsilon) + \eta, ~ \eta \sim \mathcal{N}(0,\sigma)$ to emulate detector noise.

\section{Results and Discussion}
\label{sec:results}
We will demonstrate our approach using a proof-of-concept 2D toy example with 2D meta-structures (invariant along $z$ dimension). 
For simplicity, we will assume that the intensity map $u_\text{2D}$ is uniform. 
We will also assume effectively-monochromatic (but incoherent) illumination so that the wavelength dependence may be suppressed throughout. 
Without loss of generality, we will also ignore the detector noise $\eta$, 
assuming a strong illumination and high signal-to-noise ratio. 
We will lift all these assumptions and simplifications in future works, 
which will rigorously emulate realistic scenes and lighting conditions as well as consider full-3D, large-aperture volume metaoptics. 

We imagine the 1D depth map $h(x)$ as a superposition of $M$ spatial harmonics,
\begin{equation}
	h(x)=\sum_{m=1}^Ma_m \sin (2\pi f_m x),\quad x\in(x_l,x_r),
\end{equation}
on which incoherent light sources are positioned. By observing the incoherent image $v$ of $h$, we are tasked to \textit{infer} the parameters $a_m,f_m$. It is important to note that this so-called “period-finding” problem is well known to be highly nonlinear (and an important step in the celebrated Shor’s algorithm) unlike the linear Fourier Transform which tries to estimate just the $a_m$'s at known frequencies. Our goal, then, is to design a meta-structure such that the resulting image $v$ facilitates the retrieval of both the amplitudes $a_m$ and the spatial frequencies $f_m$. In our approach, we will combine the metaoptics frontend with a light-weight \textit{linear} processing backend to allow negative processing of positive intensity values. We emphasize that nonlinear operations are provided by the metaoptics frontend, not by the computational backend, in contrast to current co-designed metaoptics-inference systems~\cite{lin2021end,lin2022end}, where all the nonlinear processing occurs in the backend.

Numerically, we uniformly sample $x$ to generate $N$ incoherent point sources and randomly construct a training dataset comprising $N_{\rm train}$ samples. 
For each sample, the finite difference frequency domain (FDFD) method is used to simulate the image $v$. 
A subsequent linear mapping, $Av+b$, where $A$ is a matrix and $b$ is a bias vector, 
is applied to predict the amplitudes and frequencies. 
We adopt the mini-batch Adam optimization algorithm to jointly design the structure and determine $A$ and $b$
so as to minimize the mean squared relative error (MSRE) loss function.
An additional $N_{\rm test}$ randomly generated test samples are employed to evaluate the inversion performance of the proposed platform.

\begin{figure*}[htbp]
	\centering
	\includegraphics[scale=0.36]{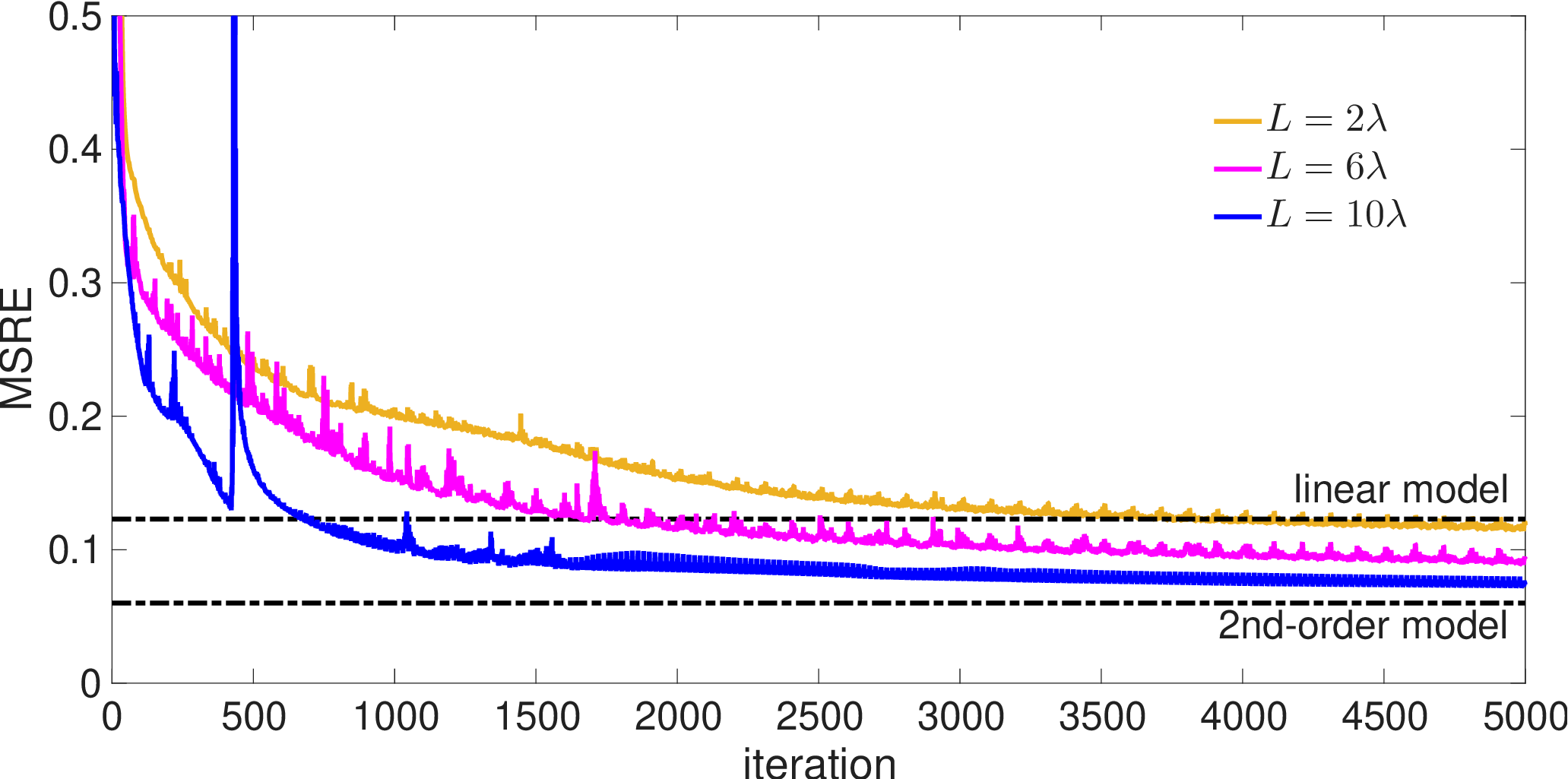}
	\caption{MSRE on testing data versus iteration number for various heights $H$.}\label{result}
\end{figure*}

\begin{figure*}[htbp]
	\centering
	\includegraphics[scale=0.36]{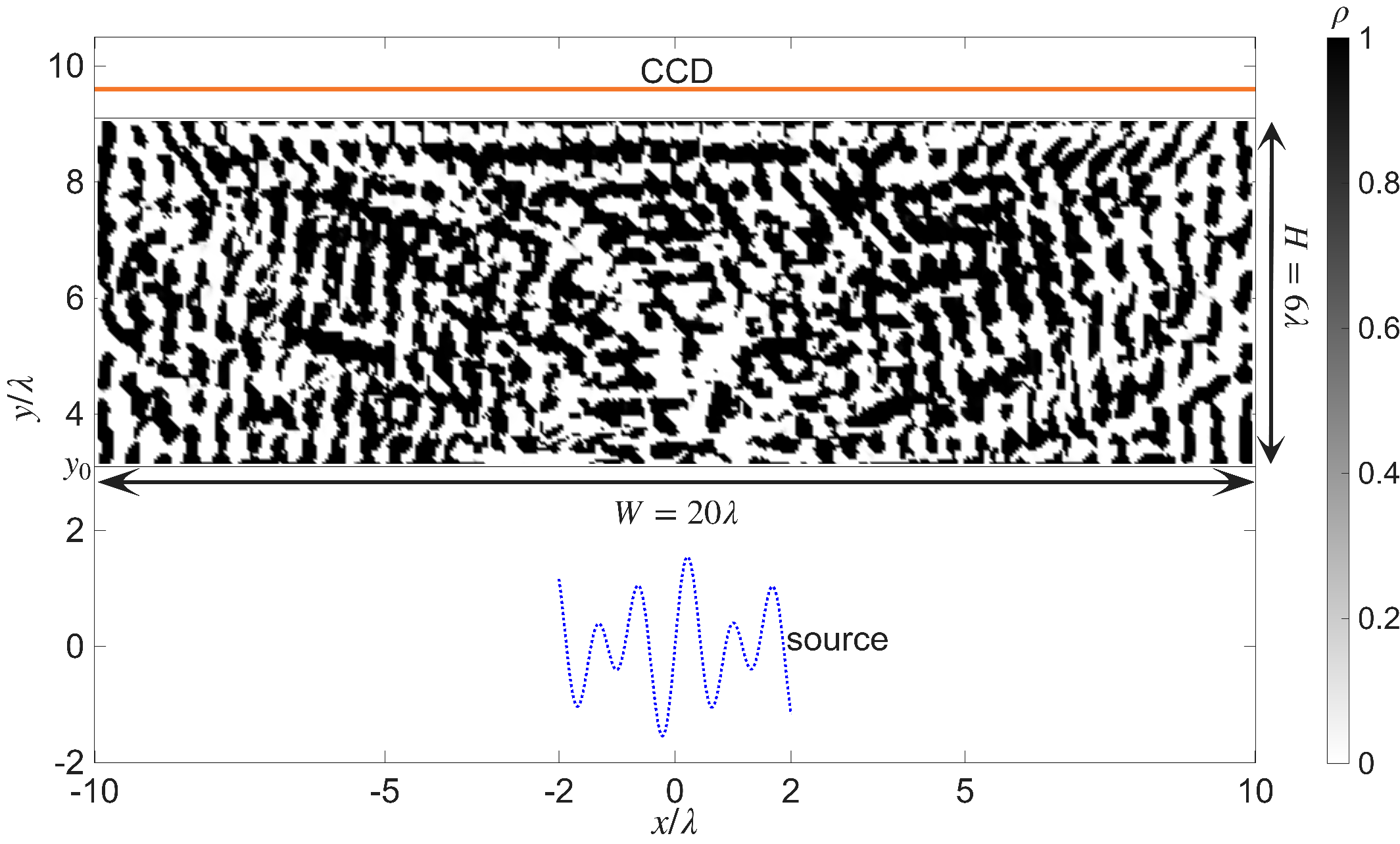}
	\caption{Optimized meta-structure for $H=6\lambda$. 
		The gray‐value map indicates the structural density $\rho$.}\label{metaoptics}
\end{figure*}

We consider a specific configuration with $x_l=-2\lambda$, $x_r=-x_l$, $M=2$, $a_m\in(0.5,1)$, $f_m\in(0.5,3)$, and $N=101$,
The dataset comprises $N_{\rm train}=50,000$ training samples and $N_{\rm test}=10,000$ testing samples. 
The meta‐optical structure occupies the spatial domain $(-W/2,W/2)\times(y_0,y_0+H)$ with its width fixed at $W=20 \lambda$.
A mini-batch size of 250 was used, and the model was trained for 20 epochs.
Figure \ref{result} presents the MSRE evaluated on the test set. 
It is evident that increasing the metasurface height $H$ enhances test accuracy, 
implying that a larger volume provides greater degrees of freedom and thus yields improved performance.
This conclusion similarly holds when $H$ is held constant and $W$ is increased.
Our results further illustrate that the platform serves as an optical analogue of a neural network:
relative increases in both its width and depth substantially improve the model’s accuracy.
In addition, in Fig.~\ref{metaoptics}, we show the optimized structure for $H=6\lambda$ as a representative example.
Here, the structural density $\rho$ parametrizes the local permittivity via $\varepsilon=\varepsilon_b+\rho(\varepsilon_{\tiny \mbox{SiO}_2}-\varepsilon_b)$
with the background permittivity set to $\varepsilon_b=1$.

Note that the black dashed line in Fig.~\ref{result} marks the 12\% MSRE achieved by retrieving amplitudes and frequencies via a linear least‐squares fit. 
Our numerical examples show that for $H=2\lambda$,
the platform reaches this $12\%$ threshold after 4,000 iterations. 
Increasing the height to $H=6\lambda$ 
allows the design to fully surpass the linear model, and at $H=10\lambda$ 
we attain an MSRE of $6.7\%$ after 5,000 iterations.  
These results indicate that incoherent imaging inherently exploits nonlinear features, 
thereby realizing effective nonlinear transforms within a linear optics.

{\it Model limitations:} We note that, although the image $v$
in our incoherent-imaging model depends nonlinearly on the depth map $y=h(x)$,
this nonlinearity appears exclusively as pure powers $y^n$, $n=1,2,3,\cdots$, 
with no mixed cross-terms of the form $y^p{y'}^q$, where $y'$ denotes the depth value at a different spatial location.
This decoupling arises from the statistical independence of incoherent sources, 
which precludes any inter-source interactions. 
Numerical results show that a meta-optics platform of height $H=10\lambda$
achieves an MSRE of 6.7\%, 
essentially matching the performance of a second-order least-squares inversion augmented with quadratic coordinate terms. 
Further increasing the platform’s height and width could drive inversion accuracy beyond the second-order regime, 
and tuning the optimization hyperparameters may yield additional gains, but the performance will remain fundamentally limited by the absence of the cross-multiplication terms. We also note the rather close proximity and limited field of view in which the sample has to be situated for maximal sensitivity to depth, corresponding to an input numerical aperture (NA) of 0.891 (= $n\sin\theta$). At this NA, even when the metaoptics is scaled to a larger aperture ($\sim$1cm), the sample distance will be constrained to a relatively close $\sim$2.55 mm.







\section{Generalization and Outlook}
\label{sec:outlook}
While our proof-of-concept 2D example may appear niche and contrived, we note that many non-trivial 3D scenes in the real world are effectively opaque. For example, the human face is 3D and opaque, suggesting that it may be possible to develop \textit{almost-}all-optical face recognition systems (with significantly reduced digital backends). More importantly, however, \textit{almost-}all-optical nonlinear vision may be extended, beyond opaque 3D scenes, to many types of complex light-matter interactions in the real world. It is important to realize that the scene intensity function $u(x,y,z,\lambda)$ in conventional image-formation models is a rather ``superficial description'' of an illuminated scene---a computationally-convenient representation that suppresses complex, \textit{latent} information which otherwise would have been \textit{perceived} by a highly-sensitive metaoptics. At a more fundamental level, any real-world scene or object is characterized by a \textit{latent} permittivity $\varepsilon_\text{scene}$ and permeability $\mu_\text{scene}$ distribution. When illuminated (coherently, incoherently or even partially-coherently), these distributions can give rise to multiple scatterings of waves. These complex interactions are then borne out by outgoing wave signals, to be captured by an imaging optics. 

One could even create a scenario where multiple back-and-forth scatterings are introduced and reinforced between the scene $\varepsilon_\text{scene}, \mu_\text{scene}$ and the metaoptics $\varepsilon_\text{metaoptics}$. Such a setup could allow the final image to acquire an arbitrarily complex nonlinear dependence on $\varepsilon_\text{scene}, \mu_\text{scene}$---offering a potential universal function approximator that can be trained on freeform metaoptics permittivity $\varepsilon_\text{metaoptics}$ and even perhaps complex illumination patterns with \textit{dynamically} tunable spectrum, polarization and coherence properties. Since $\varepsilon_\text{scene}, \mu_\text{scene}$ is the most fundamental and accurate description of nature (sans the quantum regime), being able to directly sense and process these latent distributions, using highly complex volume metaoptics, could potentially unlock novel meta-human ``latent-vision'' capabilities, as opposed to conventional anthropocentric RGB vision based on refractives, diffractives or even single-layer metasurface optics. 

Looking forward, one critical challenge to realizing all-optical deep \textit{latent} vision would be the expensive computational and experimental costs associated with designing and fabricating complex volume metaoptics as well as the costs associated with full-wave ab-initio rendering of real-world scenes. We note that 3D nano-fabrication is rapidly maturing with unprecedented resolutions and throughputs~\cite{liang2025two,dorrah2025free}. Therefore, we argue that it may also be worthwhile to invest considerable effort into developing new, ultra-scalable computational ecosystems~\cite{lin2019overlapping,sun2025scalable,xue2023fullwave,lin2022fast,skarda2022low,mahlau2025flexible,lu2025systolic} for rapidly simulating and optimizing large, complex electromagnetic arenas ($\varepsilon_\text{scene}, \mu_\text{scene}, \varepsilon_\text{metaoptics}$). Fortuitously, such efforts will also dovetail nicely with the frontier of AI research on generative world models and high-entropy virtual training environments~\cite{abu2023interactive}, which will address the scarcity of high-quality training data for deep \textit{unsupervised} learning beyond traditional RGB vision.

\section*{Acknowledgement}
This work is supported by the US Army Research Office (ARO) under Contract No. W911NF2410390.

\bibliographystyle{unsrt}
\bibliography{ndlvm}
\end{document}